\documentclass[pra,twocolumn,superscriptaddress]{revtex4}
\usepackage{bm,graphicx,amsmath}
\usepackage{placeins}
\usepackage{amssymb}
\usepackage{amsmath}
\usepackage{pstricks}
\usepackage{epsfig}
\usepackage{dsfont}
\usepackage{upgreek}
\usepackage{tipa}
\newcommand{\ket}[1]{\left| #1 \right\rangle}
\newcommand{\bra}[1]{\left\langle #1 \right|}

\newcommand{\avrg}[1]{\left\langle #1\right\rangle}

\newcommand{\ie}{{\it{i.e.}}}

\newcommand{\mE}{{\mathcal E}}
\begin{document}
\title{
Two-color quantum memory in double $\Lambda$-media
}

\date{\today}

\author{D. Viscor}
\affiliation{Departament de F\'{\i}sica, Universitat Aut\`{o}noma de Barcelona, E-08193 Bellaterra, Spain} 
\author{V. Ahufinger}
\affiliation{Departament de F\'{\i}sica, Universitat Aut\`{o}noma de Barcelona, E-08193 Bellaterra, Spain} 
\author{J. Mompart}
\affiliation{Departament de F\'{\i}sica, Universitat Aut\`{o}noma de Barcelona, E-08193 Bellaterra, Spain} 
\author{A. Zavatta}
\affiliation{European Laboratory for Nonlinear Spectroscopy (LENS), Via Nello Carrara 1, I-50019 Sesto Fiorentino (Florence), Italy}
\affiliation{Istituto Nazionale di Ottica, INO-CNR, L.go E. Fermi, 6, I-50125, Florence, Italy}
\author{G. C. La Rocca}
\affiliation{Scuola Normale Superiore and CNISM, Piazza dei Cavalieri, 56126 Pisa, Italy} 
\author{M. Artoni}
\affiliation{European Laboratory for Nonlinear Spectroscopy (LENS), Via Nello Carrara 1, I-50019 Sesto Fiorentino (Florence), Italy}
\affiliation{Department of Physics and Chemistry of Materials CNR-IDASC Sensor Lab, Brescia University, Brescia, Italy} 

\begin{abstract}

We propose a quantum memory for a single-photon wave packet in a superposition of two different colors, \ie, two different frequency components, using the electromagnetically induced transparency technique in a double-$\Lambda$ system. We examine a specific configuration in which the two frequency components are able to exchange energy through a four-wave mixing process as they propagate, so the state of the incident photon is recovered periodically at certain positions in the medium. We investigate the propagation dynamics as a function of the relative phase between the coupling beams and the input single-photon frequency components. Moreover, by considering time-dependent coupling beams, we numerically simulate the storage and retrieval of a two-frequency-component single-photon qubit.
\end{abstract}

\maketitle

\section{Introduction}\label{intro}

In recent years optical quantum memories had become the focus of an important research activity \cite{QM_review,Zhao'09,Riedmatten'10,Reim'10,Saglamyurek'11,Clausen'11,LongdellNews'11} for being one of the main ingredients for quantum information processing applications. 
In particular, in quantum information, the long distance transmission of photons (or flying qubits), which are the preferred information carriers, is limited by photon losses. 
Thus, transporting quantum states of light between different nodes of a quantum network requires the use of quantum repeaters \cite{qnet,qrep}, whose basic components are quantum memories. Therefore, quantum memories should be capable of storing arbitrary quantum states of light for an arbitrarily long time and release them on demand and with high efficiency and fidelity \cite{QM_review}.

Among the different methods for implementing a quantum memory, the approach based on electromagnetically induced transparency (EIT) \cite{Fleischhauer'00,Fleischhauer'02,Lukin'03,Phillips'01,Liu'01,Longdell'05} is one of the most used, allowing to store a single photon in solid state systems for times $>1$ s \cite{Longdell'05}. This technique consists in slowing down a weak light pulse coupled to one transition of a $\Lambda$-type three-level system in the presence of a control field coupled to the other optical allowed transition. By adiabatically turning off the control field the light pulse is absorbed and mapped into the coherence between the ground states. Next, after a desired time which should be smaller than the decay time of the ground states coherence, the control field is turned on again and the initial light pulse is recovered.


For the storage of a general photonic qubit, \ie, a single photon in an arbitrary superposition of two different components, more sophisticated schemes are needed \cite{Polycarpou'12,Garcia-Maraver'04,polmem_ensembles,Specht'11,Carreno'10,ViscorV'11,ViscorL'11,Camacho'09,Raczynski'02,Raczynski'04,Phillips'11,Li'05}. 
In quantum communication with photons, the logical qubits can be encoded in several ways, for example, via polarization, time bin, path, phase, photon number or even frequency encodings. 
In particular, several works have focused on the storage of photons with two frequency components using the EIT technique in resonant double-$\Lambda$ media \cite{Lin'09,Wu-Stationary'10,Wu-Decay'10,Wu'09,Camacho'09,Raczynski'02,Raczynski'04,Phillips'11,Li'05}. Those proposals have been formulated mainly in the semiclassical regime. 
However, the storage of a two-color quantum entangled state would be interesting because it would have potential applications in future quantum information networks, {\it e.g.}, they could be used to link systems of different nature \cite{Lloyd'01,Shapiro'02}.
One of the main issues regarding two-color memories, both in quantum and in classical approaches, is that the existence of a dark state in resonant double-$\Lambda$ systems \cite{Liu'10,Chong'08} together with the presence of four-wave mixing processes lead to a pulse matching of the frequency components \cite{Merriam'00}. This implies that the two input frequency modes can not be independently stored \cite{Phillips'11}. In particular, only a specific combination of the two modes can be perfectly absorbed and recovered \cite{Merriam'00,Li'06,Raczynski'04}, whereas for an arbitrary two-frequency-mode input, part of the light will propagate transparently and part will be absorbed \cite{Raczynski'04}. 
Nonetheless, it has been shown that the four-wave mixing processes arising in double-$\Lambda$ media, which make difficult the implementation of a suitable quantum memory, have interesting applications in frequency conversion of classical probe beams \cite{Korsunsky'99,Merriam'00}, 
single-photon frequency conversion preserving the quantum coherence \cite{Chong'08}, 
and in the possibility to combine or redistribute one or two previously stored frequency modes \cite{Li'06,Li'05,Raczynski'02}, even with different relative intensities \cite{Raczynski'04}. 
An interesting situation is found when one of the two $\Lambda$ systems of the double-$\Lambda$ configuration is far detuned from the one photon resonance. In this case, considering the continuous wave regime, it has been shown that the total light intensity is weakly absorbed during the propagation \cite{Jain'96,Korsunsky'99}, while the intensity of each mode oscillates sinusoidally with the optical length, being the energy transferred back and forth between the two probe beams. Later, a similar result was obtained in the quantum regime \cite{Payne'03}, where a single photon coupled initially to one of the transitions of the double-$\Lambda$ system oscillates during propagation between the two frequency modes, thus creating a superposition state at certain positions in the medium with high efficiency.
 
In this work, we combine the usual EIT-based storage technique with the four-wave mixing properties of a double-$\Lambda$ system to implement a quantum memory for single photons in an arbitrary superposition state of two frequencies. By solving the evolution equations of the single-photon frequency components we show that an arbitrary input superposition of two frequency modes can be recovered at certain positions of the medium, and that the relative phase between the coupling fields and the particular form of the input state play a crucial role in the propagation dynamics of the frequency components. Later, the storage and retrieval of the frequency superposition state is shown by the numerical integration of the system equations.

The paper is organized as follows. 
In Sec.~\ref{sec:TheModel} we describe the physical system and derive the equations that govern its evolution.
In Sec.~\ref{sec:Solutions} we analytically solve the propagation equations of the incident single-photon frequency components and study several examples using different input superposition states and control fields parameters.
Next, in Sec.~\ref{sec:Numerics}, numerical integration of the evolution equations of the system is performed to check the validity of the analytical approach, and the storage and retrieval of a particular input superposition state is presented.
Finally, we summarize the results of this work and present the main conclusions in Sec.~\ref{sec:Conclusions}.

\section{The model}
\label{sec:TheModel}

\begin{figure}
\includegraphics[width=0.8\columnwidth]{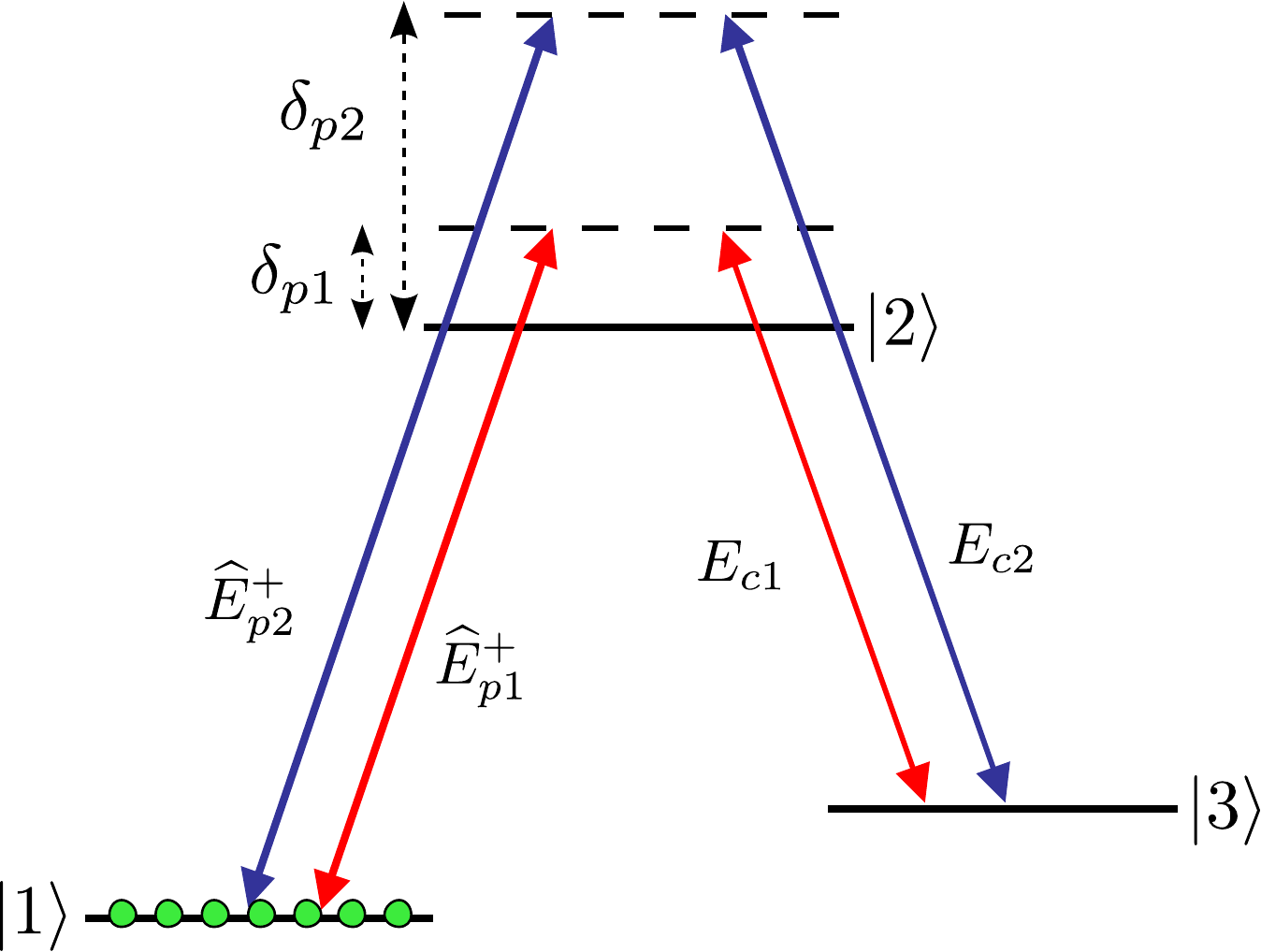}
\caption{(color online). Double-$\Lambda$ atomic scheme coupled with single-photon frequency components, $\widehat{E}_{p1}^{+}$ and $\widehat{E}_{p2}^{+}$, and classical field amplitudes, $E_{c1}$ and $E_{c2}$, satisfying the two-photon resonance condition. The fields are detuned from the one photon resonance with corresponding detunings $\delta_{p1}$ and $\delta_{p2}$, with $\delta_{p1}\ll\delta_{p2}$. All the atoms are initially in state $\left|1\right\rangle$.}
\label{f:scheme}
\end{figure}

We consider the physical system sketched in Fig.~\ref{f:scheme}, where a single-photon wave packet in a superposition of two different frequency modes, of central frequencies $\omega^{0}_{p1}$ and $\omega^{0}_{p2}$, and corresponding amplitudes $\widehat{E}_{p1}^{+}$ and $\widehat{E}_{p2}^{+}$, propagates through a system formed by $\Lambda$-type three-level atoms. 
Both frequency components interact with the left optical transition of the three-level atoms with a different detuning, $\delta_{p1}$ or $\delta_{p2}$, being $\delta_{p2}$ far from the one photon resonance while $\delta_{p1}$ is close to resonance. We assume that the difference between the detunings is much larger than the spectral widths of the frequency components, such that there is no overlap between them. The other optical transition is driven by two strong coupling beams, of frequencies $\omega^{0}_{c1}$ and $\omega^{0}_{c2}$, tuned in two-photon resonance with the corresponding single-photon components, thus forming a double-$\Lambda$ system. We consider that initially all the atoms are in the ground state $\left|1\right\rangle$. The total decay rate by spontaneous emission from the excited to the ground states is $\gamma_{2}$, and the decoherence rate of the ground states is denoted by $\gamma_{13}$.
%
%
The total Hamiltonian of the system is given by three contributions, the atomic ($H_{A}$), the field ($H_{F}$), and the interaction ($H_{I}$) Hamiltonians
\begin{eqnarray} \label{Hamiltonian}
H_{A} & = & \hbar\sum_{i=1}^{N}(\omega_{1}\hat{\sigma}_{11}+\omega_{2}\hat{\sigma}_{22}+\omega_{3}\hat{\sigma}_{33}), \\
H_{F} & = & \int_{-\infty}^{\infty}\hbar\omega_{p}\hat{a}_{\omega_{p}}^{\dagger}\hat{a}_{\omega_{p}}d\omega_{p}, \\
H_{I} & = & -\sum_{i=1}^{N}\left(\mu_{12}\hat{\sigma}_{21}^{(1)}\widehat{E}_{p1}^{+}+\mu_{12}\hat{\sigma}_{21}^{(2)}\widehat{E}_{p2}^{+}\right. \notag \\
&&\left.+\hat{\sigma}_{23}^{(1)}\hbar\Omega^{0}_{c1}+\hat{\sigma}_{23}^{(2)}\hbar\Omega^{0}_{c2}+H.c.\right),
\end{eqnarray}
where the sums are over all the $N$ atoms of the medium, the atomic population and coherence operators are of the form $\hat{\sigma}_{\nu\nu}=\left|\nu\right\rangle \left\langle\nu\right|$ and $\hat{\sigma}_{\nu\rho}^{(j)}=\left|\nu\right\rangle \left\langle\rho\right|^{(j)}$, respectively, where $\nu\neq\rho=\left\{1,2,3\right\}$ and $j=\left\{1,2\right\}$ refers to the coherence generated by the mode $\omega^{0}_{pj}$. The energy of the atomic state $\left|\nu\right\rangle$ is given by $\hbar\omega_{\nu}$, $\hbar$ being the Planck's constant, $\mu_{12}$ is the electric dipole moment of the $\left|1\right\rangle-\left|2\right\rangle$ transition, and $\hat{a}_{\omega_{p}}^{\dagger}$ and $\hat{a}_{\omega_{p}}$ are the creation and annihilation field operators, respectively, for a frequency mode $\omega_{p}$. The Rabi frequencies of the classical beams are denoted by $\Omega^{0}_{cj}=\Omega_{cj}e^{-i\omega^{0}_{cj}\left(t-z/c\right)+i\phi_{j}}$, where $\Omega_{cj}=\left|\mu_{23}\right|\left|E_{cj}\right|/\hbar$, $\mu_{23}$ is the dipole moment of the $\left|3\right\rangle-\left|2\right\rangle$ transition and $E_{cj}$ the corresponding electric field amplitude. The amplitudes of the quantum field operators read
\begin{equation} \label{vacuumfields}
\widehat{E}_{pj}^{+} = \int_{-\infty}^{\infty}\epsilon_{\omega_{p}}^{(j)}\hat{a}_{\omega_{p}}e^{+i\frac{\omega_{p}}{c}z}d\omega_{p},
\end{equation}
where $c$ is the speed of light in vacuum and $\epsilon_{\omega_{p}}^{(j)}=\epsilon\sqcap\left[(\omega_{p}-\omega^{0}_{pj})/\Delta\omega_{pj}\right]$, with $\epsilon=\sqrt{\frac{\hbar \omega_{12}}{2\varepsilon_{0}V}}$, $\varepsilon_{0}$ the electric permittivity in vacuum, $\omega_{12}=\omega_{1}-\omega_{2}$ the transition frequency between states $\left|1\right\rangle$ and $\left|2\right\rangle$, $V$ the quantization volume, and $\sqcap(\omega)$ a boxcar function of width $\Delta\omega_{pj}$, centered at $\omega^{0}_{pj}$. We assume $\Delta\omega_{pj}$ much larger than the spectral width of the corresponding frequency component $\delta\omega_{pj}$, but not enough to overlap with the other one, {\it i.e.}, $\left|\omega^{0}_{p2}-\omega^{0}_{p1}\right|\gg\Delta\omega_{pj}\gg\delta\omega_{pj}$.

To find the evolution equations of the single-photon frequency components we adopt the procedure and formalism from Ref. \cite{Moiseev_approach}. The initial state of the system has the form
\begin{align} \label{initialstate}
\left|\psi_{i}(t\rightarrow-\infty)\right\rangle = & \int d\omega_{p1}f_{\omega_{p1}}^{(1)}(-\infty)\hat{a}_{\omega_{p1}}^{\dagger}|0,0\rangle_{p}|1\rangle \notag \\
& +\int d\omega_{p2}f_{\omega_{p2}}^{(2)}(-\infty)\hat{a}_{\omega_{p2}}^{\dagger}|0,0\rangle_{p}|1\rangle. 
\end{align}
where $\omega_{pj}\in[\omega^{0}_{pj}-\Delta\omega_{pj}/2,\omega^{0}_{pj}+\Delta\omega_{pj}/2]$. Here we have used the notation $|n_{1},n_{2}\rangle_{p}|\nu\rangle$, where $n_{1}$ and $n_{2}$ are the number of photons in modes $\omega^{0}_{p1}$ and $\omega^{0}_{p2}$, respectively, and $\nu$ denotes the atomic state. Tracing out the atomic part, the first and second terms in the right-hand side of Eq.~(\ref{initialstate}) correspond to the initial state of each frequency component of the single photon, and $f_{\omega_{pj}}^{(j)}(-\infty)$ are the envelope functions of the wave packet, which have a narrow peak at $\omega_{pj}^{0}$. We assume that they are spectrally separated enough such that their overlap is negligible, and must satisfy $\int d\omega_{p1}\left|f_{\omega_{p1}}^{(1)}(-\infty)\right|^{2}+\int d\omega_{p2}\left|f_{\omega_{p2}}^{(2)}(-\infty)\right|^{2}=1$. So, the photon is initially in an arbitrary superposition of the two frequency components, and the atoms are all in the ground state $|1\rangle$. 
Next, we assume that the general form of the state of the system at any time is
\begin{equation} \label{generalstate}
	|\psi(t)\rangle=|\psi_{1}(t)\rangle+|\psi_{2}(t)\rangle+|\psi_{3}(t)\rangle,
\end{equation}
where the first, the second and the third terms correspond to the excitation being either in one of the photonic modes, in the atomic state $\ket{2}$, or in state $\ket{3}$, respectively. Their explicit forms are
\begin{eqnarray} \label{generalstateterms}
	\left|\psi_{1}(t)\right\rangle & = & \int d\omega_{p1}f_{\omega_{p1}}^{(1)}(t)\hat{a}_{\omega_{p1}}^{\dagger}|0,0\rangle_{p}|1\rangle \notag \\
	&& +\int d\omega_{p2}f_{\omega_{p2}}^{(2)}(t)\hat{a}_{\omega_{p2}}^{\dagger}|0,0\rangle_{p}|1\rangle, \label{generalstatefield} \\
	\left|\psi_{2}(t)\right\rangle & = & \sum_{i=1}^{N}\left(b_{1}(t)\hat{\sigma}_{21}^{(1)}|0,0\rangle_{p}|1\rangle+b_{2}(t)\hat{\sigma}_{21}^{(2)}|0,0\rangle_{p}|1\rangle\right), \label{generalstateexcited} \notag \\ \\
\left|\psi_{3}(t)\right\rangle & = & \sum_{i=1}^{N}g(t)\hat{\sigma}_{31}|0,0\rangle_{p}|1\rangle, \label{generalstatemetas}
\end{eqnarray}
where $b_{1}(t)$ and $b_{2}(t)$ are the probability amplitudes of exciting one atom to the state $\ket{2}$ through modes $\omega^{0}_{p1}$ and $\omega^{0}_{p2}$, respectively, and $g(t)$ is the probability amplitude of transferring the population to state $\ket{3}$ via a two-photon process. Those functions together with $f_{\omega_{p1}}^{(1)}(t)$ and $f_{\omega_{p2}}^{(2)}(t)$, give a complete description of the state of the system. In order to find their evolution, we insert the general form of the state of the system [Eq.~(\ref{generalstate})] into the Schr\"odinger equation and apply $\langle1,0|_{p}\langle1|$, $\langle0,1|_{p}\langle1|$,
$\langle0,0|_{p}\langle2|$, and $\langle0,0|_{p}\langle3|$, obtaining:
\begin{eqnarray} \label{timeevoleqs}
	i\partial_{t}f_{\omega_{pj}}^{(j)}(t) & = & \omega_{pj}f_{\omega_{pj}}^{(j)}(t)-\frac{\mu_{12}}{\hbar}N\epsilon b_{j}(t)e^{-i\frac{\omega_{pj}}{c}z}, \label{timeevolfj} \\
i\partial_{t}b_{j}(t) & = & \omega_{2}b_{j}(t)-g(t)\Omega^{0}_{cj} \notag \\
	&& -\frac{\mu_{12}}{\hbar}\epsilon\int d\omega_{pj}f_{\omega_{pj}}^{(j)}(t)e^{i\frac{\omega_{pj}}{c}z}, \label{timeevolbj} \\
	i\partial_{t}g(t) & = & \omega_{3}g(t)-\left[\left(\Omega^{0}_{c1}\right)^{*}b_{1}(t)+\left(\Omega^{0}_{c2}\right)^{*}b_{2}(t)\right]. \label{timeevolg} \notag \\
\end{eqnarray}
Next, multiplying Eq.~(\ref{timeevolfj}) by $e^{i\omega_{pj}z/c}$ and integrating over $\omega_{pj}$ we obtain the propagation equation for the quantum field amplitudes
\begin{equation} \label{evoleqfield}
\left(\frac{1}{c}\partial_{t}+\partial_{z}\right)\mE_{j}(z,t)=i\kappa_{12}\beta_{j}(z,t),
\end{equation}
where we have defined $\mE_{j}(z,t)e^{-i\omega_{pj}^{0}\left(t-z/c\right)}=\frac{\mu_{12}}{\hbar}\epsilon \int d\omega_{pj}f_{\omega_{pj}}^{(j)}(t)e^{i\omega_{pj}z/c}$, $\beta_{j}(z,t)=b_{j}(t)e^{i\omega_{pj}^{0}\left(t-z/c\right)}$, and $\kappa_{12}=\frac{N\left|\mu_{12}\right|^{2}\epsilon^{2}}{\hbar^{2}c}$. With these definitions, Eqs.~(\ref{timeevolbj}) and (\ref{timeevolg}) read
\begin{eqnarray}
\partial_{t}\beta_{j}(z,t) &=& i\mbox{\ensuremath{\Delta_{pj}}}\beta_{j}(z,t)+i\mE_{j}(z,t)+ig(z,t)\overline{\Omega}_{cj}, \label{evoleqoptcoh} \\
\partial_{t}g(z,t) &=& i\left[\overline{\Omega}_{c1}^{*}\beta_{1}(z,t)+\overline{\Omega}_{c2}^{*}\beta_{2}(z,t)\right]-\gamma_{13}g(z,t), \label{evoleqspincoh} \notag \\
\end{eqnarray}
where we have added phenomenologically the ground-states decoherence $\gamma_{13}$ in Eq.~(\ref{evoleqspincoh}) and the spontaneous emission from the excited level $\gamma_{2}$ in Eq.~(\ref{evoleqoptcoh}) through the complex detuning $\Delta_{pj}=\delta_{pj}-i\gamma_{2}/2$, being $\delta_{pj}=\omega_{pj}^{0}-\omega_{2}$. Moreover, we have defined $\overline{\Omega}_{cj}=\Omega_{cj}e^{i\phi_{j}}$, we have assumed degenerate ground states, {\it i.e.}, $\omega_{1}=\omega_{3}$ and $\omega_{pj}^{0}=\omega_{cj}^{0}$, and we have chosen the energy origin at $\hbar\omega_{1}=0$.

\section{Solutions of the evolution equations}
\label{sec:Solutions}

The equations describing the propagation of the single-photon frequency components in the double-$\Lambda$ system can be solved by using the adiabatic approximation for Eq.~(\ref{evoleqoptcoh}), {\it i.e.}, $\partial_{t}\beta_{j}(z,t)\simeq0$, and changing from temporal to frequency domain by applying the Fourier transform to our system equations. Next, inserting Eqs.~(\ref{evoleqoptcoh}) and (\ref{evoleqspincoh}) into the Fourier transformed Eq.~(\ref{evoleqfield}), a linear system of partial differential equations for the quantum field amplitudes in the frequency domain, $\widetilde{\mE}_{j}(z,\omega)$, is obtained. This can be solved, leading to
\begin{align}\label{FrequencySolution}
	\widetilde{\mE}_{j}(z,\omega) =\ & \widetilde{\mE}_{j}(0,\omega)\frac{\left|\Omega_{cj}\right|^{2}}{\left|\Omega\right|^{2}}\left(e^{i\omega z/v_{a}}+\frac{\left|\Omega_{cl}\right|^{2}}{\left|\Omega_{cj}\right|^{2}}e^{i\omega z/v_{b}}e^{i\alpha z}\right)\nonumber \\
	& + \widetilde{\mE}_{l}(0,\omega)\frac{\overline{\Omega}_{cj}\overline{\Omega}_{cl}^{*}}{\left|\Omega\right|^{2}} \left(e^{i\omega z/v_{a}}-e^{i\omega z/v_{b}}e^{i\alpha z}\right)
\end{align}
where $j,l=\left\{1,2\right\}$ and $j\neq l$, $\widetilde{\mE}_{j}(0,\omega)$ is the boundary condition for the spectral envelope of the frequency component centered at $\omega^{0}_{pj}$, $\alpha=-\kappa_{12}\frac{\left|\Omega\right|^{2}}{D}$ and 
\begin{eqnarray} \label{groupveloc}
	\frac{1}{v_{a}}&=&\frac{1}{c}+\frac{\kappa_{12}}{\left|\Omega\right|^2}, \label{groupveloca} \\
	\frac{1}{v_{b}}&=&\frac{1}{c}+ \frac{\kappa_{12}}{\left|\Omega\right|^2}\frac{\left|\Omega_{c1}\right|^2\left|\Omega_{c2}\right|^2\left(\Delta_{p1}-\Delta_{p2}\right)^2}{D^2}, \label{groupvelocb}
\end{eqnarray}
with $D=\Delta_{p1}\left|\Omega_{c2}\right|^2+\Delta_{p2}\left|\Omega_{c1}\right|^2$ and $\left|\Omega\right|^2=\left|\Omega_{c1}\right|^2+\left|\Omega_{c2}\right|^2$. 
In Eq.~(\ref{FrequencySolution}), we have assumed that the decoherence time of the ground states is much larger than the time needed to store and retrieve the single photon, thus $\gamma_{13}\simeq0$. Moreover, we have approximated the exponents as linear functions of $\omega$ by Taylor expansion up to first order, and we have considered the coefficients independent of $\omega$, as done in Ref. \cite{Payne'03}. With these approximations, and considering a small decay from the excited level $\gamma_{2}\ll(\delta_{p1}\left|\Omega_{c2}\right|^2+\delta_{p2}\left|\Omega_{c1}\right|^2)/\left|\Omega\right|^2$, the inverse Fourier transform of the field can be performed analytically. 
Then, it can be seen that in general each of the components of the frequency superposition will split in two different parts, each one propagating with a different velocity given by Eqs.~(\ref{groupveloca}) and (\ref{groupvelocb}). 
%
%
Assuming $\left|\Omega_{c1}\right|=\left|\Omega_{c2}\right|$, $\delta_{p1}=0$, and
$\delta_{p2}\gg\gamma_{2}$, the velocities for the frequency components are approximately equal $v_{b}\simeq v_{a}\equiv v$, and hence the inverse Fourier transform of Eq.~(\ref{FrequencySolution}) can be rewritten as
\begin{eqnarray} \label{Fieldsolaprox}
\mE_{j}(z,t) & = & \frac{1}{2}\left[\mE_{j}\left(0,t-\frac{z}{v}\right)\left(1+e^{i\alpha z}\right)\right. \notag \\
&& \left.+\mE_{l}\left(0,t-\frac{z}{v}\right)e^{i\phi_{jl}}\left(1-e^{i\alpha z}\right)\right],
\end{eqnarray}
where $\phi_{jl}\equiv\phi_{j}-\phi_{l}$ is the phase difference between the coupling fields. Note that $\alpha$ now reduces to $\alpha\simeq-2\frac{\kappa_{12}}{\delta_{p2}}+i\kappa_{12}\frac{2\gamma_{2}}{\delta^{2}_{p2}}$ \cite{Note-on-refr}.

To obtain the intensity of each component of the single-photon frequency superposition we calculate $\mE_{j}(z,t)\mE^{*}_{j}(z,t)$ using Eq.~(\ref{Fieldsolaprox}):
\begin{widetext}
\begin{eqnarray} \label{ProbAmp}
\left|\mE_{j}(z,t)\right|^{2} & = & \frac{1}{4}\left[ \left|\mE^{0}_{j}\left(t-\frac{z}{v}\right)\right|^{2}\left\{1+e^{-2{\rm Im}(\alpha)z}+2\cos\left[{\rm Re}(\alpha)z\right]e^{-{\rm Im}(\alpha)z}\right\}+\right.\nonumber \\
 & & +\left|\mE^{0}_{l}\left(t-\frac{z}{v}\right)\right|^{2}\left\{1+e^{-2{\rm Im}(\alpha)z}-2\cos\left[{\rm Re}(\alpha)z\right]e^{-{\rm Im}(\alpha)z}\right\}+\nonumber \\
 & & \left.+2{\rm Re}\left(\left|\mE^{0}_{l}\left(t-\frac{z}{v}\right)\right|\left|\mE^{0}_{j}\left(t-\frac{z}{v}\right)\right|e^{i\varphi_{jl}}e^{i\phi_{jl}}\left\{1-e^{-2{\rm Im}(\alpha)z}+2i\sin\left[{\rm Re}(\alpha)z\right]e^{-{\rm Im}(\alpha)z}\right\}\right)\right],
\end{eqnarray}
where $\varphi_{jl}=\varphi_{j}-\varphi_{l}$ is the phase difference between the single-photon frequency components at the input $z=0$. In Eq.~(\ref{ProbAmp}) the time evolution appears only in the boundary conditions $\mE^{0}_{j}\left(t-\frac{z}{v}\right)\equiv\mE_{j}(0,t-\frac{z}{v})$. This means that the single-photon wave packet keeps its shape, but it is drifted in time a quantity $t_{c}=z/v$, which depends on the velocity $v$ defined in Eq.~(\ref{groupveloca}). Moreover, it can be seen that, while the single photon propagates, 
the intensities of the two frequency components exhibit complementary oscillations with a rate that depends on ${\rm Re}(\alpha)$. 
Note that the decaying terms ${\rm Im}(\alpha)$ in Eq.~(\ref{ProbAmp}) are due to spontaneous emission from the excited level. 
An interesting case is found when one considers a symmetric superposition state at the input, \ie, $\left|\mE^{0}_{j}\left(t\right)\right|=\left|\mE^{0}_{l}\left(t\right)\right|$. In this situation, Eq.~(\ref{ProbAmp}) takes the form
\begin{eqnarray} \label{ProbAmpSym}
	\left|\mE_{j}(z,t)\right|^{2} & = & \left|\mE^{0}_{j}\left(t-\frac{z}{v}\right)\right|^{2}\left( \frac{1+e^{-2{\rm Im}(\alpha)z}}{2}\right.\nonumber \\
	& & \left.+\left\{\cos\left(\phi_{jl}+\varphi_{jl}\right)\left(\frac{1-e^{-2{\rm Im}(\alpha)z}}{2}\right)-\sin\left(\phi_{jl}+\varphi_{jl}\right)\sin\left[{\rm Re}(\alpha)z\right]e^{-{\rm Im}(\alpha)z}\right\}\right).
\end{eqnarray}
In this case, when $\phi_{jl}+\varphi_{jl}=0$, there is no oscillation between the frequency components during the propagation and the intensity of each single-photon frequency component is perfectly transmitted, \ie, $\left|\mE_{j}(z,t)\right|^2=\left|\mE^{0}_{j}\left(t-\frac{z}{v}\right)\right|^2$. This can be interpreted by considering that, through a four-wave mixing process mediated by the coupling beams, the energy going from the first to the second component is compensated by the energy transfer from the second component to the first one.

Analogously, we find the relative phase between the two frequency components using
\begin{eqnarray} \label{Phase}
\mE_{j}(z,t)\mE_{l}^{*}(z,t) & = & \left(\left|\mE^{0}_{j}\left(t-\frac{z}{v}\right)\right|\left|\mE^{0}_{l}\left(t-\frac{z}{v}\right)\right|\left\{\cos\left(\varphi_{jl}-\phi_{jl}\right)\frac{1+e^{-2{\rm Im}(\alpha)z}}{2}+i\sin\left(\varphi_{jl}-\phi_{jl}\right)e^{-{\rm Im}(\alpha)z}\cos\left[{\rm Re}(\alpha)z\right]\right\}\right.\nonumber \\
 & & +i\sin\left[{\rm Re}(\alpha)z\right]e^{-{\rm Im}(\alpha)z}\frac{\left|\mE^{0}_{j}\left(t-\frac{z}{v}\right)\right|^{2}-\left|\mE^{0}_{l}\left(t-\frac{z}{v}\right)\right|^{2}}{2}\nonumber \\
 & & \left.+\frac{1-e^{-2{\rm Im}(\alpha)z}}{2}\frac{\left|\mE^{0}_{j}\left(t-\frac{z}{v}\right)\right|^{2}+\left|\mE^{0}_{l}\left(t-\frac{z}{v}\right)\right|^{2}}{2}\right)e^{i\phi_{jl}}.
\end{eqnarray}
\end{widetext}
From this expression, we observe that in general the phase between the two components, $\arg{\left[\mE_{j}(z,t)\mE^{*}_{l}(z,t)\right]}$, will oscillate in a more involved way than the intensity, [Eq.~(\ref{ProbAmp})]. In particular, we observe that only when the imaginary part of the outermost parentheses in Eq.~(\ref{Phase}) vanishes the phase will be independent of $z$. 

\begin{figure}
{\includegraphics[width=\columnwidth]{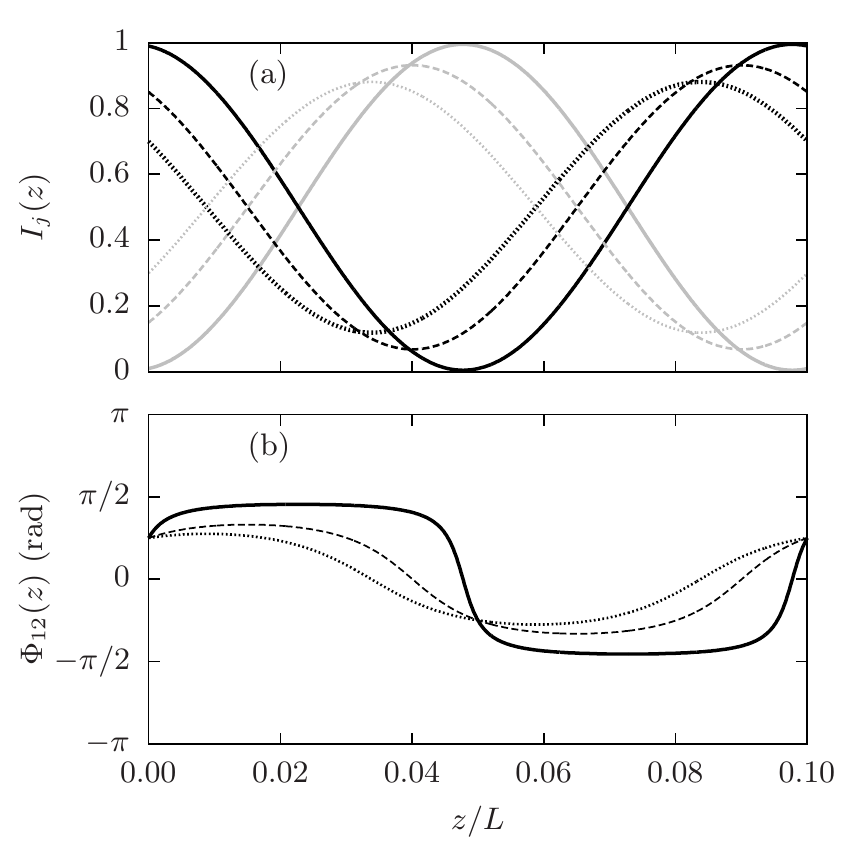}}
\caption{(a) Normalized intensities of the single-photon frequency components $I_{1}$ (black) and $I_{2}$ (gray), and (b) the relative phase between them for ${\rm Re}(\alpha)=2\pi/L$, with $L=0.1$ (a.u.) being the length of the medium, and ${\rm Im}(\alpha)=0$.
Solid lines: $I_{1}(0) = 0.99$, $\phi_{12} = 0$, $\varphi_{12} = \pi/4$; 
dashed lines: $I_{1}(0) = 0.85$, $\phi_{12} = 0$, $\varphi_{12} = \pi/4$; 
dotted lines: $I_{1}(0) = 0.70$, $\phi_{12} = 0$, $\varphi_{12} = \pi/4$.}
\label{AnalPropB}
\end{figure}

\begin{figure} 
{\includegraphics[width=\columnwidth]{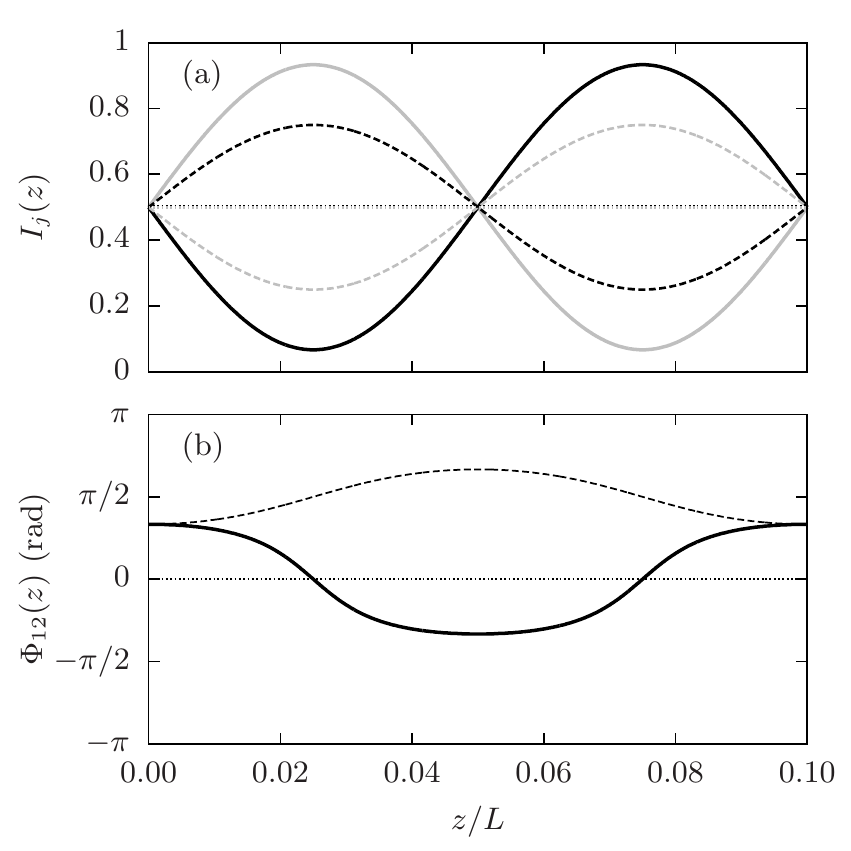}}
\caption{(a) Normalized intensities of the single-photon frequency components $I_{1}$ (black) and $I_{2}$ (gray), and (b) the relative phase between them for ${\rm Re}(\alpha)=2\pi/L$, with $L=0.1$ (a.u.) being the length of the medium, and ${\rm Im}(\alpha)=0$.
Solid lines: $I_{1}(0) = 0.5$, $\phi_{12} = 0$, $\varphi_{12} = \pi/3$; 
dashed lines: $I_{1}(0) = 0.5$, $\phi_{12} = \pi/2$, $\varphi_{12} = \pi/3$; 
dotted lines: $I_{1}(0) = 0.5$, $\phi_{12} = 0$, $\varphi_{12} = 0$.}
\label{AnalPropA}
\end{figure}

In what follows, by evaluating the analytical expressions obtained in Eqs.~(\ref{ProbAmp}) and (\ref{Phase}), we discuss different propagation examples of the two frequency components; see Figs.~\ref{AnalPropB} and \ref{AnalPropA}. We change to a reference frame fixed at the peak of the single-photon pulse $\left(t_{c}=z/v\right)$, so we need only to show the variation on the spatial dimension $z$. In Figs.~\ref{AnalPropB}(a) and \ref{AnalPropA}(a) we plot the normalized intensity of each of the two frequency components,
\begin{equation} \label{NormProb}
I_{j}(z)=\frac{\left|\mE_{j}(z,t_{c})\right|^{2}}{\left|\mE^{0}_{1}\left(t_{c}\right)\right|^{2}+\left|\mE^{0}_{2}\left(t_{c}\right)\right|^{2}},
\end{equation}
whereas in Figs.~\ref{AnalPropB}(b) and \ref{AnalPropA}(b), the relative phase between them,
\begin{equation} \label{NormPhase}
	\Phi_{jl}(z)=\arg{\left[\mE_{j}(z,t_{c})\mE^{*}_{l}(z,t_{c})\right]},
\end{equation}
is shown. In all the figures the different line styles correspond to different sets of parameters (see the caption), while the black and gray lines both in Figs.~\ref{AnalPropB}(a) and \ref{AnalPropA}(a) correspond to the intensity of the frequency components $\omega^{0}_{p1}$ and $\omega^{0}_{p2}$, respectively. We have taken ${\rm Im}(\alpha)=0$, a fact that is well justified from the assumption $\delta_{p2}\gg\gamma_{2}$ made in Eq.~(\ref{Fieldsolaprox}).
These figures are useful to show that the behavior of the two components during the propagation depends completely on the specific state at the entrance of the medium and the phase difference of the coupling beams. 
For instance, the different line styles in Fig.~\ref{AnalPropB} correspond to different initial intensities of the frequency components, while the relative phases between the frequency components and coupling beams are fixed. 
We observe that the different initial superposition states lead to intensity oscillations with different amplitudes and shifted by different amounts. 
Further examples are shown in Fig.~\ref{AnalPropA}, where the input intensities are equal for the two frequency modes, and the relative phases between them and between the coupling beams are changed. We observe in Fig.~\ref{AnalPropA}(a) that opposite behaviors for the intensity of a given mode are obtained just by properly changing the relative phase of the coupling beams (solid and dashed lines). 
Moreover, note that the case shown with dotted lines, {\it i.e.}, $I_{p1}(0)=I_{p2}(0)$ and $\phi_{12}=\varphi_{12} = 0$, corresponds to the situation discussed after Eq.~(\ref{ProbAmpSym}), in which the photon state does not evolve during propagation.

As a general conclusion from Figs.~\ref{AnalPropB} and \ref{AnalPropA}, we observe that the more different the intensities of the frequency components, the largest the variation in their relative phase, and vice versa. We also observe that the relative phase oscillates around the value $\phi_{ij}$. 
Moreover, the most remarkable fact is that the frequency of the oscillation, both in the intensity [Eq.~(\ref{NormProb})] and in the phase [Eq.~(\ref{NormPhase})] is determined only by ${\rm Re}(\alpha)\simeq-2\kappa_{12}/\delta_{p2}$. This means that by properly choosing the coupling parameter $\kappa_{12}$ and the detuning $\delta_{p2}$, one can recover at the output of the medium, $z=L$, the initially injected state with an ideally perfect fidelity, \ie, $\mE_{j}(L,t)=\mE^{0}_{j}(t-t_{c})$ for ${\rm Re}(\alpha)L=2\pi n$, with $n\in\mathds{Z}$. 
%
%

\section{Numerical analysis}
\label{sec:Numerics}

In this section we demonstrate the validity of the approximations made in the analytical approach by numerically integrating Eqs.~(\ref{evoleqfield})-(\ref{evoleqspincoh}). Moreover, we show the possibility of storing and retrieving a single photon in an arbitrary superposition state of two frequency components using time-dependent coupling fields. 
To simulate the pulse propagation in time and space, a bidimensional grid for each variable is created with a spacing in the $z$ dimension small enough to ensure the convergence of the results. The steps for the numerical protocol are the following: First, the temporal evolution of the medium variables is obtained from the incident ($z=0$) field components, which are assumed to have Gaussian profiles of temporal width $\tau=25$ ns and centered at $t_{c}=3.5\tau$, using a Runge-Kutta integrating method. Next, the field at the adjacent spatial point is determined with a finite difference method, using the preceding obtained values. Finally the previous steps are repeatedly performed until the whole grid is filled. 
For the medium we take a length of $L=0.1$ m, $\gamma_{2}\tau=0.16$, $\gamma_{13}\tau=1.6\times10^{-5}$, and $\kappa_{12}\tau L\simeq500$, while the detunings are $\delta_{p1}\tau=0$ and $\delta_{p2}\tau=160$.

\begin{figure} 
{\includegraphics[width=\columnwidth]{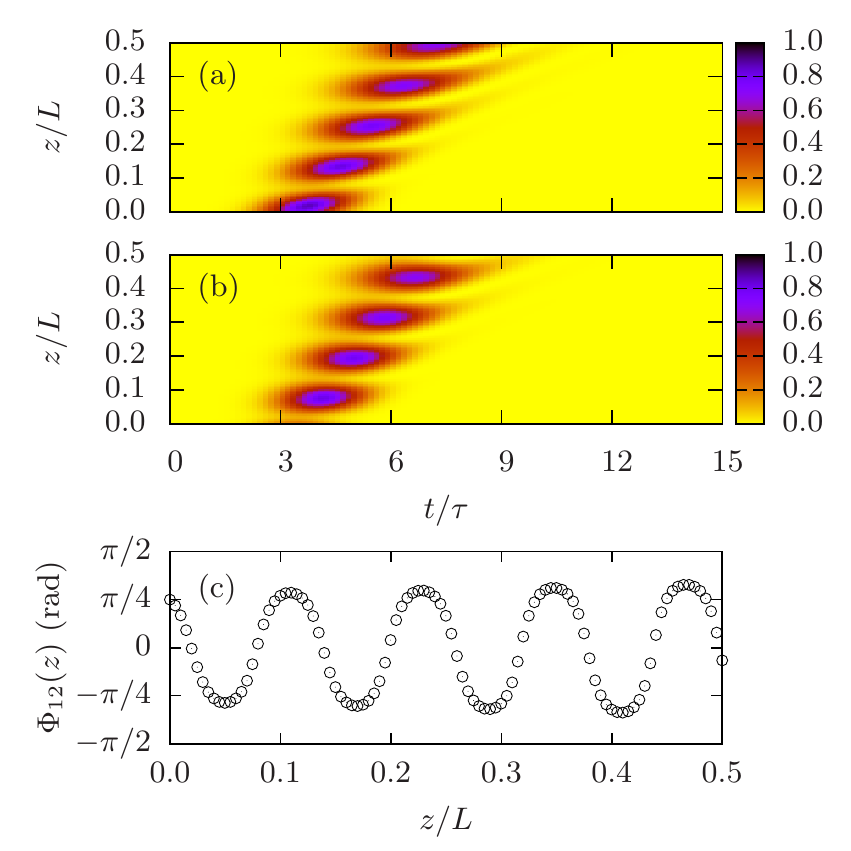}}
\caption{(color online). Normalized intensities of the single pulse frequency components, $I_{1}$ (a) and $I_{2}$ (b), as a function of normalized position and time, and (c) the phase between the frequency components at the pulse peak as a function of normalized position. The parameters correspond to the case represented with dotted lines in Fig.~\ref{AnalPropB}.}
\label{NumProp}
\end{figure}

On the one hand, an example of the propagation of the two single-photon frequency components is shown in Fig.~\ref{NumProp}, where the normalized intensity of both components, $I_{1}$ (a) and $I_{2}$ (b), is shown as a function of position and time for constant coupling Rabi frequencies $\left|\overline{\Omega}_{cj}\right|\tau=\left|\overline{\Omega}_{cl}\right|\tau=18$, and a phase between them of $\phi_{12} = 0$. The peak amplitudes of the frequency components for the injected single photon are $\left|\mE^{0}_{1}(t_{c})\right|\tau = 1.3\times10^{-3}$ and $\left|\mE^{0}_{2}(t_{c})\right|=\sqrt{0.3/0.7}\left|\mE^{0}_{1}(t_{c})\right|$, with a relative phase $\varphi_{12} = \pi/4$ between them. Note that those parameters correspond to the case represented in Fig.~\ref{AnalPropB} with dotted lines. 
As we observe, the intensities for the two frequency components exhibit complementary oscillations with a spatial period of $\sim0.125L$. Moreover, the displacement of the peak allows to estimate a propagation velocity of $\sim10^6$ m/s. 
Using the model derived in the previous section, the values obtained for the oscillation period and the velocity are $\sim0.1L$ and $\sim4.5\times10^6$ m/s, respectively. Thus, the numerical simulations are in agreement with the analytical results. 
The phase between the components at the peak of the pulse, $\Phi_{jl}(z)$, is plotted in Fig. \ref{NumProp}(c) as a function of $z$. We observe that the behavior of the phase is in good agreement with the analytical result [see dotted line in Fig.~\ref{AnalPropB}(b)].
\begin{figure}
{\includegraphics[width=\columnwidth]{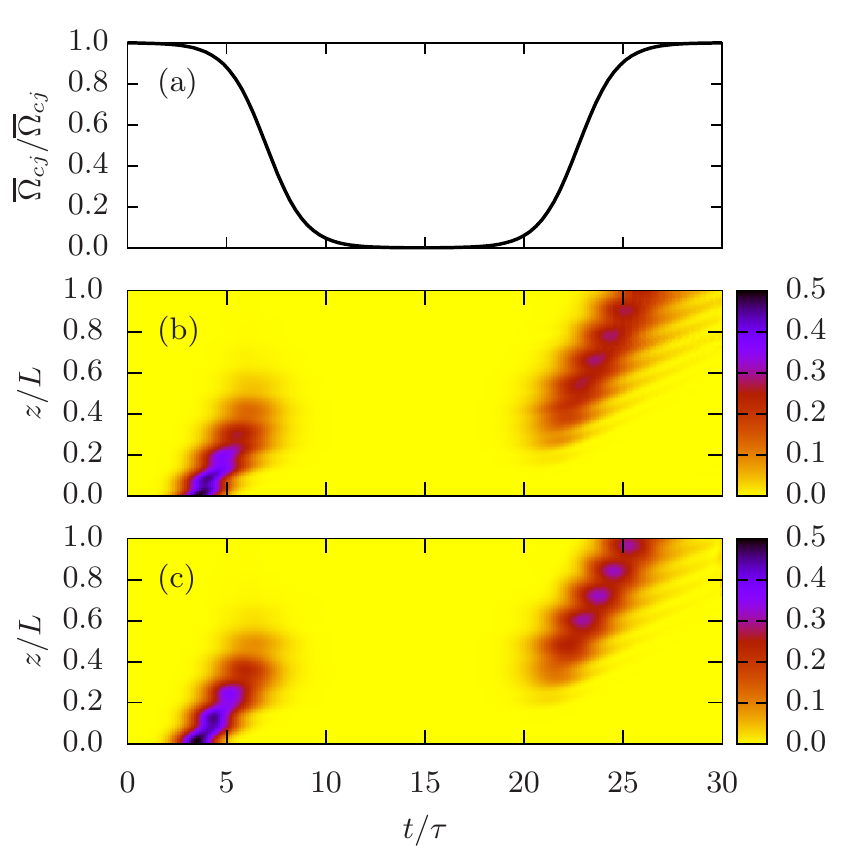}}
\caption{(color online). Temporal profile of the coupling beams (a) and normalized intensities of the single pulse frequency components, $I_{1}$ (b) and $I_{2}$ (c), as a function of normalized position and time. The parameters correspond to the case represented with dotted lines in Fig.~\ref{AnalPropA}.}
\label{NumRes}
\end{figure}

On the other hand, Fig.~\ref{NumRes} shows a particular example of the storage and retrieval process, using temporal profiles of the coupling beams of the form
\begin{equation} \label{CouplingFun}
	\overline{\Omega}'_{cj}(t) = \frac{\overline{\Omega}_{cj}}{2}\left\{2-\tanh\left[\sigma(t-t_{1})\right] + \tanh\left[\sigma(t-t_{2})\right]\right\}, \notag \\
\end{equation}
with $\left|\overline{\Omega}_{c1}\right|\tau=\left|\overline{\Omega}_{c2}\right|\tau=18$, $\sigma=0.5\tau^{-1}$, $t_{1}=2t_{c}$ and $t_{2}=6t_{c}$ [see Fig.~\ref{NumRes}(a)], and a phase difference between the coupling fields of $\phi_{12} = 0$.
In Figs.~\ref{NumRes}(b) and (c) the normalized intensity of frequency components $I_{1}$ and $I_{2}$, respectively, is shown as a function of position and time. In the example we have taken equal amplitudes for the two components of the input state, $\left|\mE^{0}_{1}(t_{c})\right|\tau =\left|\mE^{0}_{2}(t_{c})\right|\tau= 1.3\times10^{-3}$, and an initial phase difference $\varphi_{12} = 0$ between them, in such a way that the chosen parameters correspond to the situation with constant coupling fields represented with dotted lines in Fig.~\ref{AnalPropA}. 
Figure~\ref{NumRes} shows an example of how the superposition state can be stored and recovered by appropriately varying in time the coupling fields. 
Here, the storage time corresponds approximately to $t_{2}-t_{1}=0.35\,\mu$s, and it could be extended in principle to times of the order of $1/\gamma_{13}$ ($\sim1.5$ ms for the parameters considered). 
The behavior of the intensities for each component coincides with the predictions of the theoretical model. We have checked that the total pulse area is almost conserved although the pulse spreads during propagation. The phase between the frequency components (not shown in the figure) keeps an approximately constant value of $\Phi_{jl}(z)=0$ during the whole storage and retrieval process, as expected from the dotted line in Fig.~\ref{AnalPropA}(b). 
To characterize the memory performance, the efficiency of the storage and the retrieval processes, and the fidelity of the recovered superposition state have been calculated. On the one hand, we define the performance efficiency as $\eta=\eta_{\rm Abs}\eta_{\rm Ret}$, with the absorption $\eta_{\rm Abs}$ and retrieval $\eta_{\rm Ret}$ efficiencies being
\begin{eqnarray}
	\eta_{\rm Abs} &=& 1-\frac{\int_{t_{0}}^{t_{f}/2}\left(\left|\mE_{1}(L,t)\right|^2+\left|\mE_{2}(L,t)\right|^2\right)dt} {\int_{t_{0}}^{t_{f}/2}\left(\left|\mE^{0}_{1}(t)\right|^2+\left|\mE^{0}_{2}(t)\right|^2\right)dt}, \label{AbsEff} \\
	\eta_{\rm Ret} &=& \frac{\int_{t_{f}/2}^{t_{f}}\left(\left|\mE_{1}(L,t)\right|^2+\left|\mE_{2}(L,t)\right|^2\right)dt} {\int_{t_{0}}^{t_{f}/2}\left(\left|\mE^{0}_{1}(t)\right|^2+ \left|\mE^{0}_{2}(t)\right|^2\right)dt} \label{RetEff} 
\end{eqnarray}
where the interval $t_{0}=0$, $t_{f}=30\tau$ is the integration time. Computing these expressions with the data obtained in the simulation shown in Fig.~\ref{NumRes}, the absorption and retrieval efficiencies are $\eta_{\rm Abs}=99.78\%$ and $\eta_{\rm Ret}=91.21\%$, respectively. Thus, the total efficiency is $\eta=91.01\%$. 
On the other hand, the conditional fidelity is defined as $F_{c}=\left|\bra{\psi_{\rm in}}\psi_{\rm out}\rangle\right|^2$, where we take as input and output states
\begin{align}
	\ket{\psi_{\rm in}} =& \frac{\int_{t_{0}}^{t_{f}/2}\left|\mE^{0}_{1}(t)\right|^2dt} {\int_{t_{0}}^{t_{f}/2}\left(\left|\mE^{0}_{1}(t)\right|^2+ \left|\mE^{0}_{2}(t)\right|^2\right)dt}\ket{1,0}_p\ket{1} + \notag \\
	& \frac{e^{i\left\langle\varphi^{\rm in}_{12}\right\rangle}\int_{t_{0}}^{t_{f}/2}\left|\mE^{0}_{2}(t)\right|^2dt} {\int_{t_{0}}^{t_{f}/2}\left(\left|\mE^{0}_{1}(t)\right|^2+ \left|\mE^{0}_{2}(t)\right|^2\right)dt}\ket{0,1}_{p}\ket{1}, \label{phiin} \\ 
	\ket{\psi_{\rm out}} =& \frac{\int_{t_{f}/2}^{t_{f}}\left|\mE_{1}(L,t)\right|^2dt} {\int_{t_{f}/2}^{t_{f}}\left(\left|\mE_{1}(L,t)\right|^2+ \left|\mE_{2}(L,t)\right|^2\right)dt}\ket{1,0}_p\ket{1} + \notag \\
	& \frac{e^{i\left\langle\varphi^{\rm out}_{12}\right\rangle}\int_{t_{f}/2}^{t_{f}}\left|\mE_{2}(L,t)\right|^2dt} {\int_{t_{f}/2}^{t_{f}}\left(\left|\mE_{1}(L,t)\right|^2+ \left|\mE_{2}(L,t)\right|^2\right)dt}\ket{0,1}_p\ket{1} \label{phiout}
\end{align}
respectively, with $\avrg{\varphi^{\rm in}_{12}}\equiv\int_{t_{0}}^{t_{f}}\arg{\left[\mE_{1}(0,t)\mE^{*}_{2}(0,t)\right]}dt$ and $\avrg{\varphi^{\rm out}_{12}}\equiv\int_{t_{f}/2}^{t_{f}}\arg{\left[\mE_{1}(L,t)\mE^{*}_{2}(L,t)\right]}dt$. Therefore, the calculated conditional fidelity for the case shown in Fig.~\ref{NumRes} is $F_{c}=99.69\%$.

\section{Conclusions}
\label{sec:Conclusions}

In this work we have studied the propagation of a single photon, in an arbitrary superposition of two different frequency components, through a double-$\Lambda$ medium. 
In this particular configuration the intensities of the two frequency components exhibit complementary periodic oscillations as they propagate. These propagation effects have been used in combination with the light storage technique based on EIT to implement a quantum memory for frequency encoded single-photon qubits. We have studied analytically the dependence of the relative phase between the coupling fields and the input qubit state in the propagation dynamics. Moreover we have shown that, at certain positions in the medium, the initial single photon, which can be in any desired frequency superposition state of the two frequency components, is recovered. The numerical results, obtained by numerically integrating the evolution equations of the system, are in good agreement with the analytical solutions and thus the validity of the analytical approach has been confirmed. 
Finally, the storage and retrieval of a single-photon state in an arbitrary superposition of two frequency components has been shown numerically by turning off and on the coupling fields during the propagation of the single photon. For the specific choice of parameters we adopt here, the results demonstrate an efficient quantum memory for high-fidelity storage and retrieval of a frequency encoded single-photon qubit.

\begin{acknowledgments}

The authors gratefully acknowledge discussions with Alessandro Ferraro, Yury Loiko, Jin Hui Wu, and Sylwia Zielinska, and financial support through Spanish MICINN contracts FIS2008-02425, FIS2011-23719, CSD2006-00019, and HI2008-0238, the Italian Ministry MIUR through the Azione Integrata IT09L244H5, the 2011 Fondo di Ateneo of the Brescia University and the Catalan Government contract SGR2009-00347. A.Z. acknowledge support by Ente Cassa di Risparmio di Firenze, Regione Toscana, under project CTOTUS, and the EU under ERA-NET CHIST-ERA project QSCALE.

\end{acknowledgments}


\begin{thebibliography}{99}
\bibitem{Note-on-refr} It is worth noting that within this approximation the pulse loosely couples to the atomic transitions at variance with what occurs in a standard EIT configuration. The numerical analysis in Sec.~\ref{sec:Numerics} is carried out within the same limit.
\end{thebibliography}
\end{document}